**APPLICATION**

# geohabnet: An R package for mapping habitat connectivity for biosecurity and conservation

Aaron I. Plex Sulá[1,2,3*] (https://orcid.org/0000-0001-7317-3090), Krishna Keshav[1,2,3], Ashish Adhikari[1,2,3] (https://orcid.org/0000-0003-1016-0617), Romaric A. Mouafo-Tchinda[1,2,3] (https://orcid.org/0000-0003-1343-4820), Jacobo Robledo[1,2,3] (https://orcid.org/0000-0002-2707-4005), Stavan Nikhilchandra Shah[1,2,3,4] (https://orcid.org/0009-0007-7402-5912), Karen A. Garrett[1,2,3] (https://orcid.org/0000-0002-6578-1616)

[1]Plant Pathology Department, Institute of Food and Agricultural Sciences, University of Florida, Gainesville, FL, USA

[2]Emerging Pathogens Institute, University of Florida, Gainesville, FL, USA

[3]Global Food Systems Institute, University of Florida, Gainesville, FL, USA

[4]Department of Computer and Information Science and Engineering (CISE), University of Florida, Gainesville, FL, USA

*Corresponding authors: plexaaron@ufl.edu (A.I.P.S); karengarrett@ufl.edu (K.A.G.)

Current address of Romaric. A. Mouafo-Tchinda: Biopterre—Bioproducts Development Centre, Institut de Technologie Agroalimentaire, La Pocatière, Quebec, Canada

## AUTHOR CONTRIBUTIONS STATEMENT

AIPS and KAG developed the initial concepts. AIPS, KK, and SNS developed the initial version of the R package. All authors contributed to refining the R package. All authors contributed to development of the manuscript and case studies.

## ACKNOWLEDGEMENTS

We appreciate support from the USDA Animal and Plant Health Inspection Service Cooperative Agreements AP21PPQS&T00C195 and AP22PPQS&T00C133 and the USDA Specialty Crop Research Initiative, project award nos. 2020-51181-32198, 2022-51181-38242 and 2024-51181-43302, from the USDA National Institute of Food and Agriculture. We also appreciate support from the CGIAR Seed Equal Research Initiative, USAID Bureau of Humanitarian Assistance award number 720BHA22IO00136, and the CGIAR Trust Fund. We thank all donors and organizations that globally support the work of CGIAR through their contributions to the CGIAR Trust Fund. The opinions expressed in this article are those of the authors and do not necessarily reflect the views of USAID or USDA. We appreciate helpful input from Yanru Xing.

**Abstract**

1. Mapping habitat suitability, based on factors like host availability and environmental suitability, is a common approach to determining which locations are important for the spread of a species. Mapping habitat connectivity takes geographic analyses a step further, evaluating the potential roles of locations in biological invasions, pandemics, or species conservation. Locations with high habitat suitability may play a minor role in species spread if they are geographically isolated. Yet, a location with lower habitat suitability may play a major role in a species' spread if it acts as a bridge between regions that would otherwise be physically fragmented.
2. Here we introduce the geohabnet R package, which evaluates the potential importance of locations for the spread of species through habitat landscapes. Unlike most software analyzing landscape connectivity, geohabnet incorporates key factors such as dispersal probabilities and habitat availability in a network framework. These factors are often needed to better understand habitat connectivity for host-dependent species, such as pathogens, arthropod pests, or pollinators.
3. geohabnet uses publicly available or user-provided datasets, six network centrality metrics, and a user-selected geographic extent (global, national, or regional). We provide examples using geohabnet for surveillance prioritization of emerging plant pathogens and pests in Africa and the Americas. These examples illustrate how users can apply geohabnet for their species of interest and generate maps of the estimated importance of geographic locations for species spread.

**4.** geohabnet provides a quick, open-source, and reproducible baseline to quantify a species' habitat connectivity across a wide range of geographic scales and evaluates potential scenarios for the expansion of a species through habitat landscapes. geohabnet supports biosecurity programs, invasion science, and conservation biology when prioritizing management efforts for transboundary pathogens, pests, or endangered species.

## INTRODUCTION

Understanding the geographic structure of habitat networks, shaped by landscape patterns of abiotic and biotic factors, supports the conservation of endangered species. The study of habitat networks also supports invasion biology, identifying locations important for programs protecting ecosystems from the harmful effects of invasive species. Species distribution models (SDMs) are widely used in ecology to map a species' potential geographic range, but few explicitly incorporate dispersal when quantifying habitat suitability for host-dependent species such as invasive arthropod herbivores, plant pathogens, or endangered pollinators (Schmitt *et al.*, 2017, Naimi & Araújo, 2016, Velazco *et al.*, 2022). Many algorithms exist for estimating landscape connectivity, and these are commonly incorporated into biodiversity conservation planning (Brennan *et al.*, 2022, Drielsma *et al.*, 2022, Fletcher *et al.*, 2011). In landscape epidemiology and invasion science, quantifying habitat connectivity contributes to prioritizing surveillance and mitigation efforts by identifying vulnerable areas with high epidemic or invasion potential (Radici *et al.*, 2023, Jousimo *et al.*, 2014, Margosian *et al.*, 2009, Meentemeyer *et al.*, 2012). However, we are not aware of options in R for calculating habitat or landscape connectivity that integrate both species dispersal and geographic host distribution (Rimbaud *et al.*, 2024, Jones *et al.*, 2021). Both components are key to mapping the spatial spread and potential distribution of host-dependent species such as plant pathogens, arthropod parasites (Xing *et al.*, 2020, Cunniffe *et al.*, 2015), endophytes, and pollinators (Bishop, 2025).

Over 100 metrics have been proposed to measure habitat or landscape connectivity (Rayfield *et al.*, 2011, Kindlmann & Burel, 2008, Keeley *et al.*, 2021). However, few open-source resources are available for applying most of these metrics. In landscape ecology, connectivity metrics are generally based on three widely-used approaches: least-cost corridor analysis (McRae, 2006, Lewis, 2025), resistance surface modeling or electrical circuit theory (McRae *et al.*, 2008), and network-based landscape connectivity (Urban & Keitt, 2001). Over the past two decades, interest in applying network theory to invasion ecology and landscape epidemiology has been growing (Mestre & Silva, 2023, Frost *et al.*, 2019). There is a need for flexible tools that readily harness the benefits of a network-based approach and geographic information systems for landscape connectivity. Network-based tools often assess landscape connectivity at the patch level (Chubaty *et al.*, 2020, Mestre & Silva, 2023, Hesselbarth *et al.*, 2019), which may underrepresent habitat heterogeneity of realistic landscapes (Spanowicz & Jaeger, 2019).

To address this gap, we introduce the geohabnet R package, a geographically explicit network-based tool that maps landscape or habitat connectivity of both host-dependent and host-independent species. Habitat connectivity for host-dependent species often depends on the spatial distribution of suitable abiotic factors (widely addressed by SDMs), compatible host species (and other biotic factors), and natural or human-mediated opportunities for dispersal. To calculate habitat connectivity in geohabnet, a habitat network represents a set of geographic locations in a habitat landscape (nodes), and weights of connections between nodes are proportional to the probability of a species' movement between habitat locations (Fig. 2). Xing et al. (2020) provide a comprehensive description of the mathematical models (gravity models, dispersal functions, network metrics) supporting the habitat connectivity framework used in geohabnet.

The probability of species movement in geohabnet depends on habitat availability, which can be determined with SDMs based on biotic or abiotic suitability, as well as dispersal ability. A network-based approach offers powerful perspectives for movement ecology in complex, real-world landscapes, ranging from identifying nearest neighbors to determining shortest paths between physically distant regions (Garrett *et al.*, 2018, Moslonka-Lefebvre *et al.*, 2011). SDMs have been applied widely to predict species' potential distributions based on abiotic factors. The case studies provided below focus on habitat suitability defined in terms of host availability and incorporate potential dispersal, providing a first approximation for habitat connectivity. Inputs in geohabnet can also be coupled with SDMs to provide an integrated perspective on habitat connectivity.

Applying this network perspective on habitat connectivity, geohabnet allows users to ask fundamental questions, such as: how important is a focus location for spread through the habitat network, as a hub or bridge? We illustrate the use of geohabnet with three case studies, focusing on habitat connectivity for invasive or re-emerging pathogens and pests because of their importance to plant health in natural systems (Savary *et al.*, 2019, Diagne *et al.*, 2021, MacLachlan *et al.*, 2021). In these examples, geohabnet harnesses publicly available or user-supplied data for host distributions relevant to plant pathogens and pests, evaluates potential scenarios for pathogen or pest dispersal, and generates regional maps of habitat landscape connectivity for surveillance prioritization.

## FUNCTIONALITY

**Implementation.** geohabnet 2.2 provides R users with two main functions, sensitivity_analysis() and msean(), to calculate the habitat connectivity of locations in a landscape. These functions allow evaluation of habitat connectivity across a set of user-specified habitat, dispersal, and geographic parameters, and commonly used network metrics (Fig. 1). Users can install geohabnet with the following code: install.packages("geohabnet") for the stable version in CRAN or devtools::install_github("GarrettLab/HabitatConnectivity", subdir = "geohabnet") for the development version in GitHub. Installation and functions of geohabnet have been successfully tested in R version 4.3.0-4.5.1 on Windows, MacOS, and Linux operating systems, as well as High-Performance Computing settings.

Once installed, users download the parameters.yaml file in their working directory using library(geohabnet) and get_parameters() (Fig. 1). Users can manually change the values of each parameter in this file, set the newly specified values in R with set_parameters(), and run sensitivity_analysis(). In the parameters.yaml file, users can provide the directory location of the habitat suitability map, which should be in raster format.

Alternatively, msean() users provide a SpatRaster object as input and modify parameter values directly in the function. These functions have been tested using maps of habitat suitability in the form of host availability from publicly available databases like MAPSPAM (IFPRI, 2019), CROPGRIDS (Tang *et al.*, 2024), EARTHSTAT (Monfreda *et al.*, 2008), CroplandCROS (https://croplandcros.scinet.usda.gov/), and GBIF (https://www.gbif.org/), as well as custom-built by the user. The types of data for habitat suitability supported by geohabnet should range from 0 (lowest habitat suitability) to 1 (highest habitat suitability), representing host availability, environmental suitability, or a combination of habitat factors.

A third way of customizing parameters is through the geohabnet app. This app makes geohabnet functionality available in a straightforward interactive dashboard for habitat connectivity (Fig. 1). The user-friendly graphical interface (i) allows easy customizations of all parameters and sub-parameters, (ii) executes the habitat connectivity analysis, and (iii) generates downloadable visualizations and outputs. To use the app, users install it with `devtools::install_github("GarrettLab/geohabnet-UI",subdir = "Geohabnetify")`, load the package with `library(geohabnetify)`, and `run_app()` in RStudio. Upon successful execution, a web window will launch, presenting the user with access to the interface.

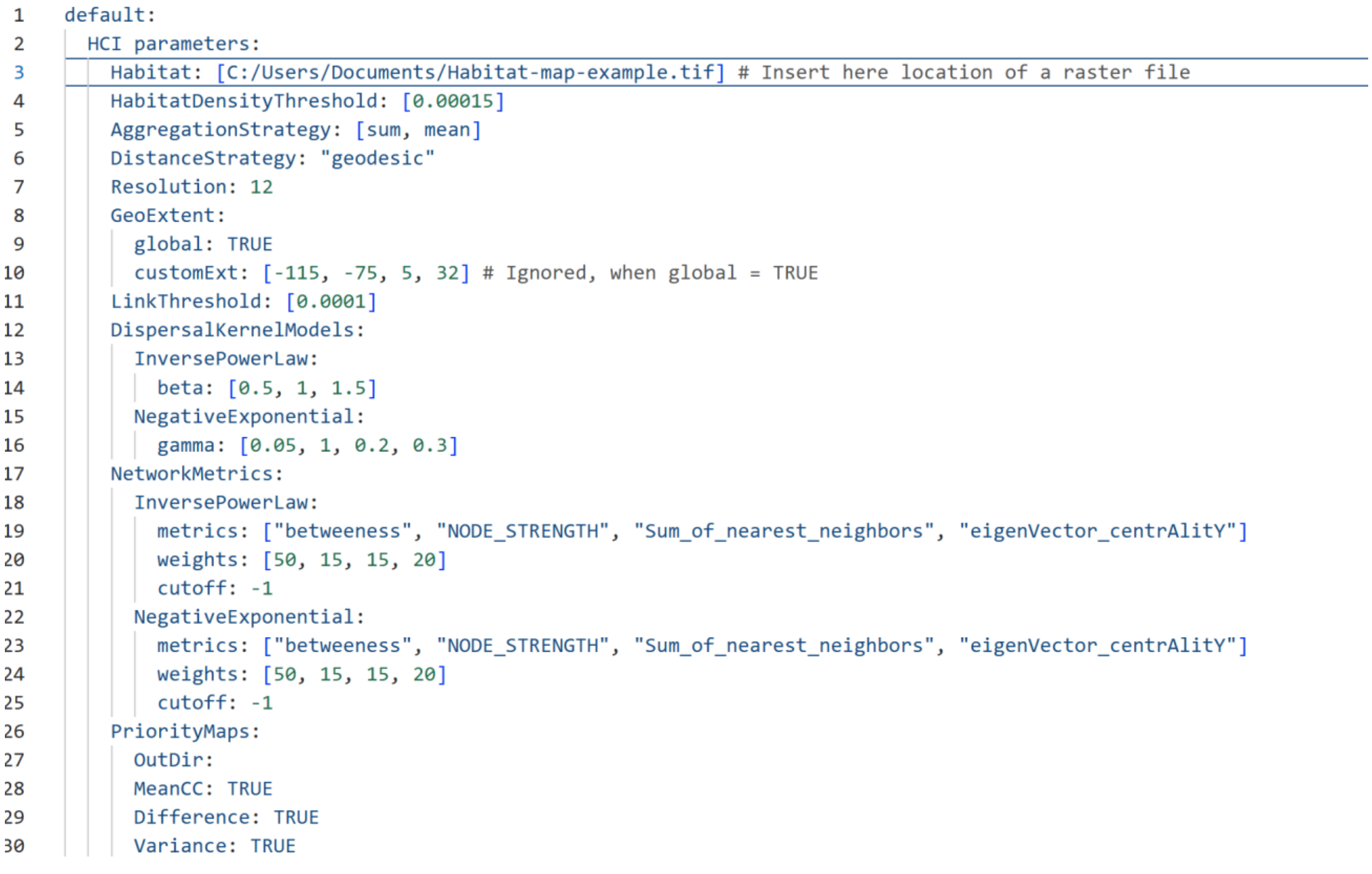

```yaml
default:
  HCI parameters:
    Habitat: [C:/Users/Documents/Habitat-map-example.tif] # Insert here location of a raster file
    HabitatDensityThreshold: [0.00015]
    AggregationStrategy: [sum, mean]
    DistanceStrategy: "geodesic"
    Resolution: 12
    GeoExtent:
      global: TRUE
      customExt: [-115, -75, 5, 32] # Ignored, when global = TRUE
    LinkThreshold: [0.0001]
    DispersalKernelModels:
      InversePowerLaw:
        beta: [0.5, 1, 1.5]
      NegativeExponential:
        gamma: [0.05, 1, 0.2, 0.3]
    NetworkMetrics:
      InversePowerLaw:
        metrics: ["betweeness", "NODE_STRENGTH", "Sum_of_nearest_neighbors", "eigenVector_centrAlitY"]
        weights: [50, 15, 15, 20]
        cutoff: -1
      NegativeExponential:
        metrics: ["betweeness", "NODE_STRENGTH", "Sum_of_nearest_neighbors", "eigenVector_centrAlitY"]
        weights: [50, 15, 15, 20]
        cutoff: -1
    PriorityMaps:
      OutDir:
      MeanCC: TRUE
      Difference: TRUE
      Variance: TRUE
```

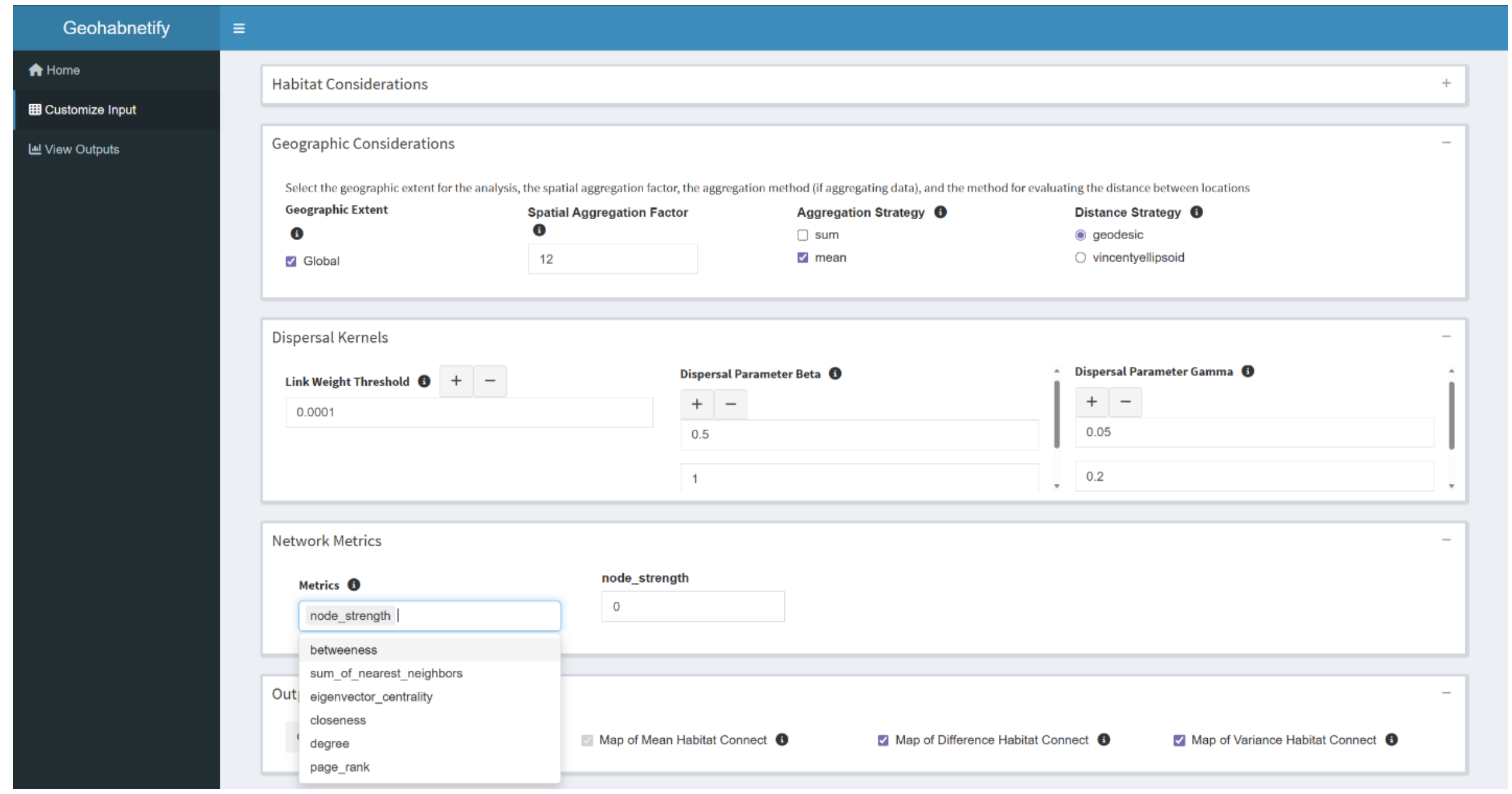

**Fig. 1.** (Top) Default values of parameters for the sensitivity_analysis() function in parameters.yaml, where parameter values can be modified by users. (Bottom) The user interface of the geohabnet app allows users to interactively choose habitat parameters.

**Outputs.** Both sensitivity_analysis() and msean() produce a map for each of the three main outputs of a habitat connectivity analysis. Both perform a sensitivity analysis by supporting evaluation of different combinations of up to 10 parameters and 24 sub-parameters (Fig. 1) in each main outcome map. The first map represents the mean habitat connectivity calculated across every combination of parameters specified by the user. This map indicates the importance of locations in the potential spread of a species in a landscape, considering the geographic pattern in habitat suitability and dispersal probabilities of the target species. The second map summarizes the variance in habitat connectivity across the same combination of specified parameters. This map helps users evaluate how habitat connectivity of a location remains consistent (no variance) or changes with different values of the model parameters. The third map indicates the difference in ranks between mean habitat connectivity and habitat suitability. This map helps identify which locations connectivity analysis indicates could allow a species to spread more readily (positive values) or less readily (negative values), compared to an analysis relying solely on habitat suitability.

Additional outputs generated by these functions include a spatRaster object for each output map and an adjacency matrix for each dispersal parameter combination. The spatRaster objects can be used for integration with other geographically explicit ecological analyses, customized visualization, or saving a copy of maps. Adjacency matrices can be used for more detailed analyses such as evaluating invasion simulations and management scenarios in the impact network analysis (INA) package (Garrett, 2021).

## CASE STUDIES

To illustrate the functionality of geohabnet, we present three case studies. A vignette for each case study is available at https://garrettlab.github.io/HabitatConnectivity/

**Case study 1**. In this case, we created a hypothetical habitat landscape with each grid cell representing a geographic location and node (Fig. 2A), similar to the example in Xing et al. (2020). This example is meant to help users better understand habitat connectivity analysis before analyzing more complex, realistic landscapes. It illustrates a simple stepwise approach to calculating the habitat connectivity of locations in a hypothetical landscape. Briefly, the habitat landscape is transformed into a species movement network (Fig. 2B), where each node represents a habitat location and link weights are proportional to the probability of species movement between locations in the landscape. The importance of each node to the spread of a species across the network is measured using a set of network centrality metrics, such as closeness centrality (Fig. 2C). The importance of each node is measured as a weighted mean of a user-selected set of network metrics. Then the mean across dispersal parameters in the sensitivity analysis is used to map the mean habitat connectivity in the landscape (Fig. 2D-G). We used the msean() function in geohabnet to easily reproduce this habitat connectivity analysis. This example illustrates how four network metrics provide different perspectives on the potential roles of locations when a species spreads in a hypothetical habitat landscape (Fig. 2D-G). Note that a location with high habitat suitability will tend to rank among the most important locations in the landscape across different network metrics. The importance of locations with low habitat suitability was higher or lower depending on the network metric used, where each metric measures a different potential role for a node in the network. We also generated an ensemble of possible spread scenarios based on a range of dispersal parameters and four network metrics, generating a mean habitat connectivity map (Fig. 2H).

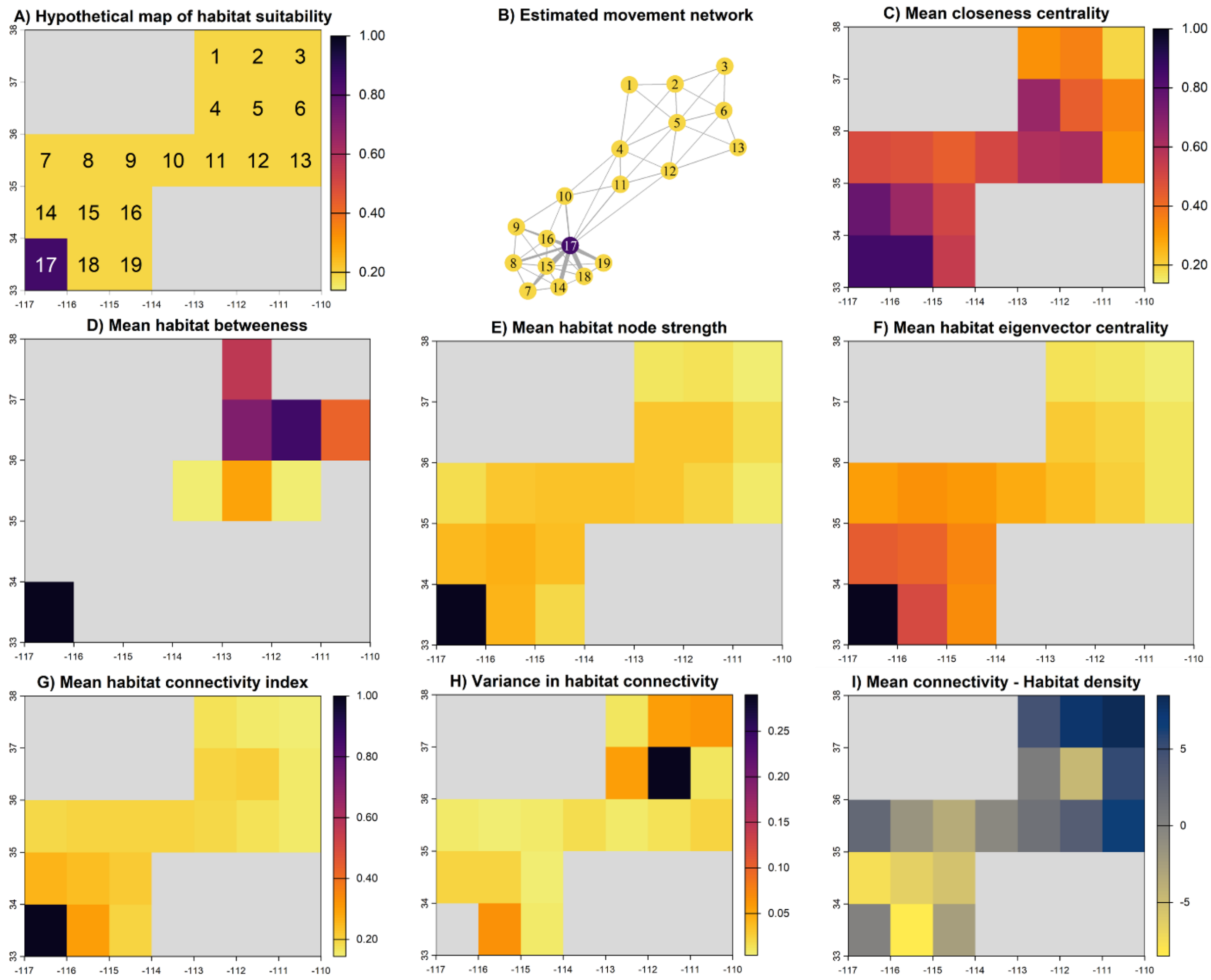

**Fig. 2.** Components of a habitat connectivity analysis in a hypothetical landscape using geohabnet. A) shows the hypothetical distribution of habitat suitability for a species, where cell 17 has unusually high suitability. B) illustrates the corresponding network of potential movement between habitat locations (nodes), with link width proportional to probability. C) to F) show the importance of habitat locations based on individual network metrics. G) shows the mean habitat connectivity across network metrics and dispersal parameters. H) is the variance in habitat connectivity across network metrics

and dispersal parameters. I) is the difference in ranks between mean habitat connectivity (G) and habitat availability (A).

**Case study 2.** This case study illustrates use of geohabnet for mapping habitat connectivity at a national, regional, or continental scale. For the threats posed by emerging crop pathogens and pests, host availability is an important factor in habitat suitability. Cropland connectivity can be used as a proxy for potential invasive spread of host-specific pathogens and pests. Here, geohabnet helped identify candidate priority locations for regional surveillance of potential invasions of pathogens and pests affecting sorghum in sub-Saharan Africa and yam in West Africa.

We first downloaded geographic data on cropland harvested area for sorghum and yam in sub-Saharan Africa from MAPSPAM 2017 (IFPRI, 2019), using the crop_spam() function in the geodata package. We used the resulting cropland density map in the msean() function in geohabnet to perform a sensitivity analysis based on cropland connectivity. We set link_threshold = 0.00001 and hd_threshold = 0.0025 (Xing et al., 2020). We customized the default weights of the four network centrality metrics in Fig. 1A: betweenness centrality (weighted 50%), node strength (weighted 15%), sum of nearest neighbors (weighted 15%), and eigenvector centrality (weighted 20%).

For sorghum in Africa, high cropland connectivity was identified in locations in southern Niger, northern Nigeria, and northern Cameroon. For yam in West Africa, the analysis identifies locations in south-eastern Nigeria and near Lake Kossou (Ivory Coast) as highly connected.

Along with information on pests and pathogens in these locations, this analysis can inform strategies for pathogen and pest surveillance in a region. For example, the spotted stem borer (*Chilo partellus*), a damaging pest of sorghum, is now present in Cameroon and prevalent in several countries in East and South Africa (van den Berge, 2023). The analysis indicates locations along the border between northern Cameroon and Nigeria as candidate priorities for mitigation to prevent the introduction of this pest into sorghum and other cereal production in West Africa.

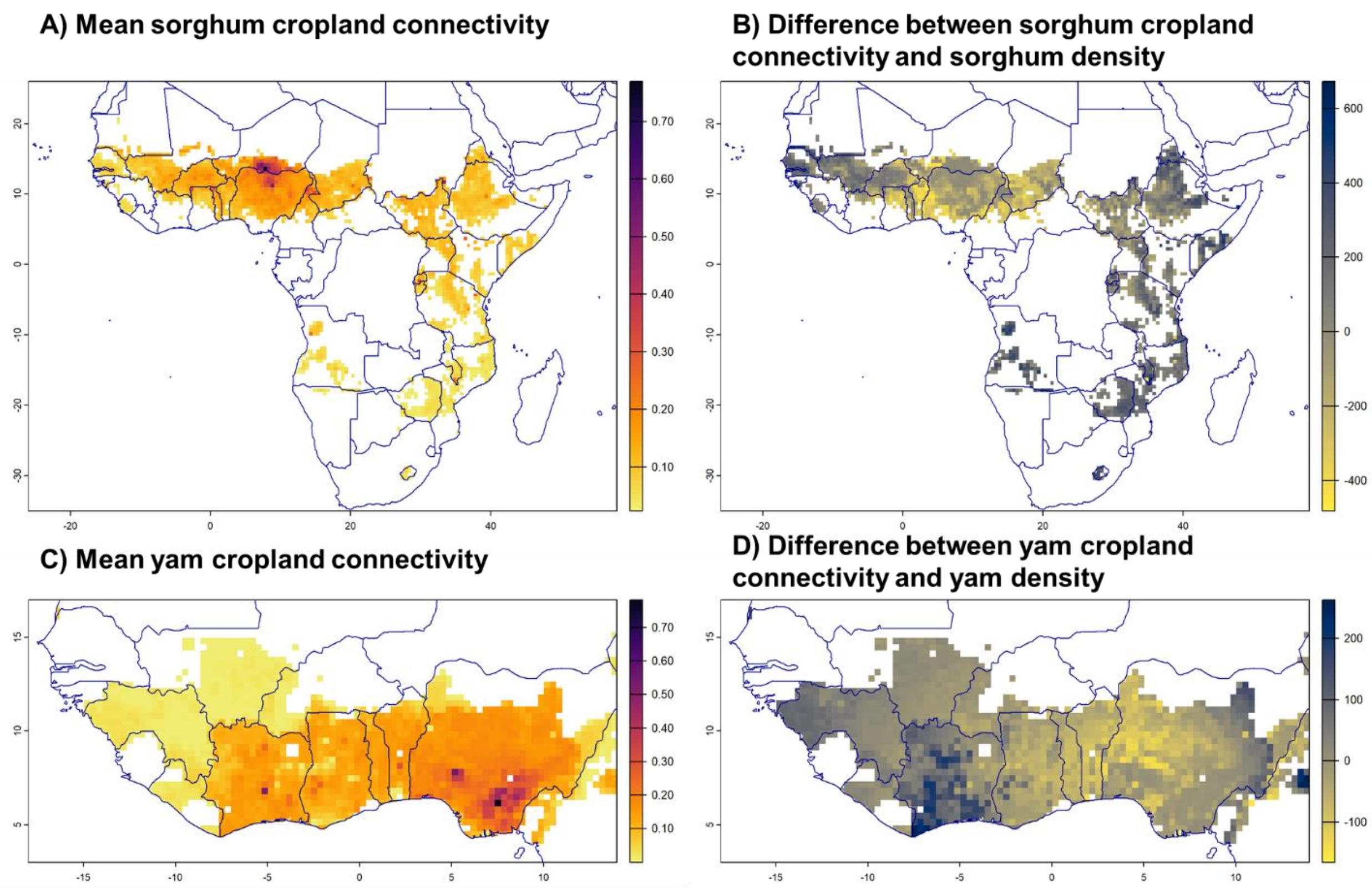

**Fig. 3.** Habitat connectivity analysis based on host availability for pathogens and pests of sorghum in sub-Saharan Africa (A, B) and of yam in West Africa (C, D).

**Case study 3.** This case study illustrates habitat connectivity analysis where habitat is widely available across a region, specifically maize in the Americas. We downloaded global maps of “maize” and “maize forage” from CROPGRIDS (Tang et al., 2024) and combined these layers to produce a cumulative map of host availability. We used this global map of maize availability in the parameters.yaml for sensitivity_analysis(). We set the host density threshold to 0.25%, link weight threshold to 0.000001, and spatial resolution to 12 degrees. To illustrate cropland connectivity resulting from potential long-distance dispersal, we estimated dispersal probabilities between locations using the inverse power law model. In a sensitivity analysis, we assessed three values of the dispersal parameter ($\beta$ = 0.5, 1, and 1.5).

There is a high mean maize cropland connectivity in major production areas where maize is densely cultivated, including the northern United States, central Mexico and southwestern Brazil (Fig. 4A). Maize is less densely cultivated in the Andes, but cropland connectivity analyses indicated locations in the Andes that were relatively more important based on cropland connectivity than based solely on maize cropland density (Fig. 4B). A region from Mesoamerica to Panama and the Andes may serve as a bridge connecting major production areas in the mainland Americas and potentially facilitating the spread of maize-specific diseases.

This map of maize connectivity in the Americas can help in monitoring the gradual spread of invasive or emerging maize-specific pests. An example of emerging diseases is maize lethal necrosis (MLN), caused in maize by a co-infection of *Maize chlorotic mottle virus* and Potyviruses. MLN has been reported with scattered occurrences in the Americas since 1974, but its recent emergence in Africa and Asia in 2011 makes this disease a global concern for maize production (Boddupalli *et al.*, 2020).

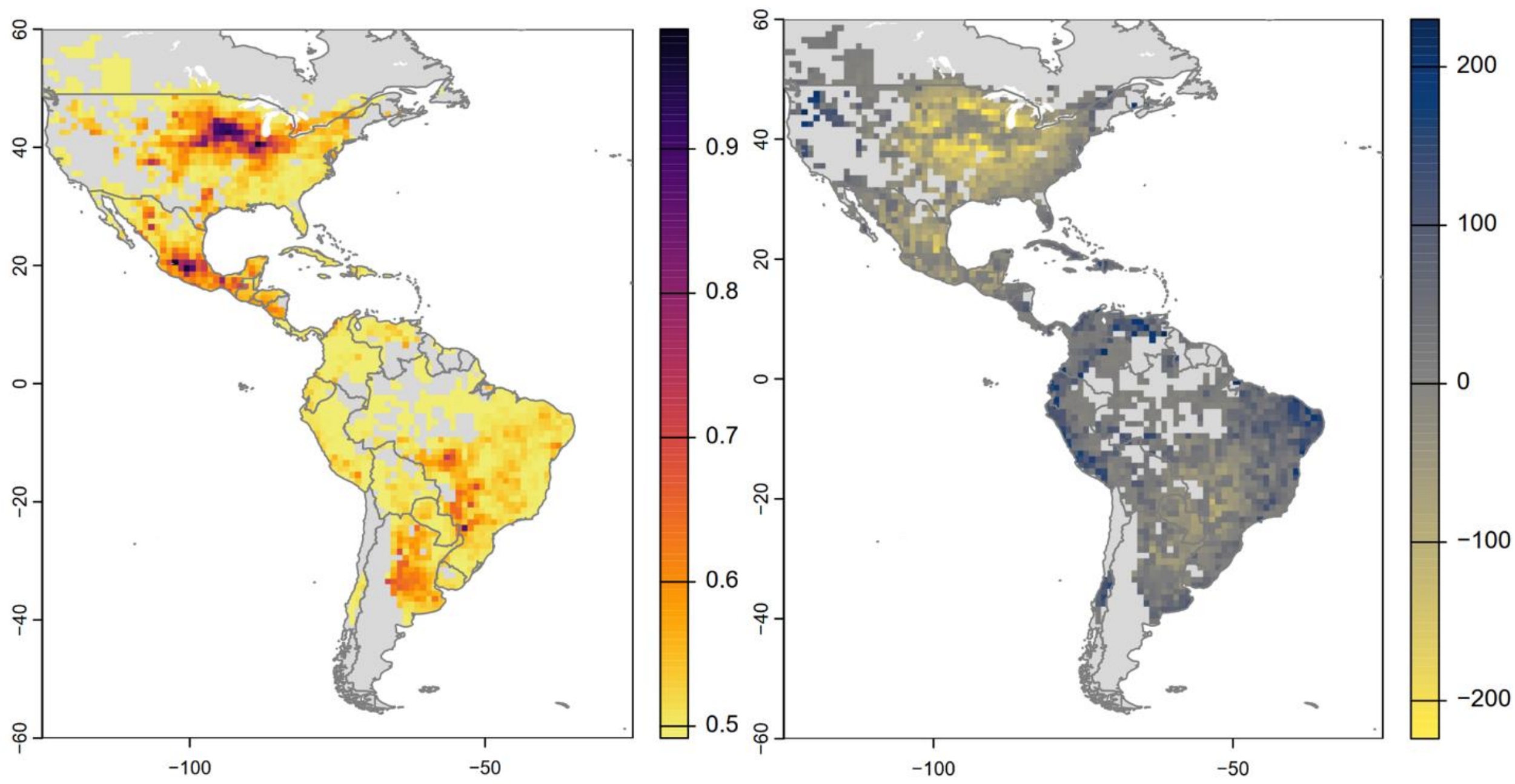

**Fig. 4.** (A) Mean maize cropland connectivity in the Americas across a range of dispersal parameter values. (B) The difference in ranks for mean maize cropland connectivity and maize density, indicating where habitat connectivity identifies locations that are more or less important compared to simply identifying locations with high maize density.

## DISCUSSION

The geohabnet package is a component of the R2M Plant Health Toolbox (https://www.garrettlab.com/r2m/), which aims to provide rapid geographic risk assessments for surveillance and mitigation of plant pathogen and pest impacts. Case studies 2-3 illustrate how geohabnet results can inform surveillance of emerging epidemics or invasive spread of crop-specific pests. In these case studies, host connectivity is not specific to one pest species because estimates of dispersal parameters for individual species are often unavailable. Habitat connectivity based on host availability is thus a first approximation to understanding the potential spread of new pests. When estimates of the dispersal parameters for a target species are unknown, stakeholders can evaluate a range of possible scenarios for dispersal parameters in a sensitivity analysis, as in case studies 2 and 3.

A more complete analysis of habitat connectivity can be constructed by incorporating, for example, geographic information about wild hosts, abiotic factors such as climate variables, and management interventions in the landscape. geohabnet has the potential for integration with a wide range of available SDMs (Velazco et al., 2022, Schmitt et al., 2017). Future versions of geohabnet may incorporate features such as habitat variability within locations. Effective strategies for species management would benefit from analysis of how habitat connectivity changes in a year or over decades, harnessing the growing availability of geographic biodiversity information (Meyer *et al.*, 2015).

## CONFLICT OF INTEREST STATEMENT

The authors declare that they have no conflicts of interest.

## DATA AVAILABILITY STATEMENT

geohabnet is available in CRAN (https://cran.r-project.org/web/packages/geohabnet/index.html) and GitHub (https://github.com/GarrettLab/HabitatConnectivity)

The user guide for geohabnet is available at https://garrettlab.github.io/HabitatConnectivity/articles/user_guide.html